\documentclass[a4paper]{jpconf}
\usepackage{graphicx}
\begin{document}
\title{Magnetic and dielectric properties of $A_2$CoSi$_2$O$_7$ ($A$=Ca, Sr, Ba) crystals}

\author{M. Akaki, J. Tozawa, D. Akahoshi, and H. Kuwahara}

\address{Department of Physics, Sophia University, Tokyo 102-8554, Japan}

\ead{m-akaki@sophia.ac.jp}

\begin{abstract}
We have investigated the magnetic and dielectric properties of $A_2$CoSi$_2$O$_7$ ($A$=Ca, Sr, and Ba) crystals with a two-dimensional network of CoO$_4$ and SiO$_4$ tetrahedra connected with each other through the corners. In Ca$_2$CoSi$_2$O$_7$, a weak ferromagnetic transition occurs at 5.7 K, where the dielectric constant parallel to the $c$ axis shows a concomitant anomaly. The large magnetocapacitance effect is observed below 5.7 K; $\Delta \varepsilon (H)/\varepsilon (0) \equiv [\varepsilon (H)-\varepsilon (0)]/\varepsilon (0)$ reaches 13 \% at 5.1 K\@. These results indicate a strong coupling between the magnetism and dielectricity in Ca$_2$CoSi$_2$O$_7$. Sr$_2$CoSi$_2$O$_7$ shows a similar magnetoelectric behavior to that of Ca$_2$CoSi$_2$O$_7$. In contrast, in Ba$_2$CoSi$_2$O$_7$, which has the different arrangement of SiO$_4$ and CoO$_4$ tetrahedra from that of Ca$_2$CoSi$_2$O$_7$, the magnetocapacitance is hardly observed. The key for the magnetocapacitance effect of $A_2$CoSi$_2$O$_7$ lies in the quasi-two-dimensional crystal structure.
\end{abstract}

\section{Introduction}
Since the discovery of the giant magnetoelectric effect in TbMnO$_3$ \cite{TbMn}, multiferroic materials that are both magnetic and dielectric have been attracting much attention. The mechanism of the magnetic ferroelectricity is well explained in terms of a spin-current model proposed by Katsura {\it et al} \cite{Katsura}. According to this model, the ferroelectricity of multiferroic materials originates in a spiral spin structure. Therefore, materials with spin frustration and/or nontrivial spin structures have attracted renewed interest as promising candidates for new magnetoelectrics. 
 
In this context, we expect $A_2$CoSi$_2$O$_7$ ($A$=Ca, Sr, and Ba) as one of such candidates because $A_2$CoSi$_2$O$_7$ is a derivative of Ba$_2$CuGe$_2$O$_7$, which has a spiral spin structure below 3.26 K \cite{BaCu}. However, the magnetic transition temperature of Ba$_2$CuGe$_2$O$_7$ is rather low probably because of a spin fluctuation of Cu$^{2+}$ ($S$=1/2), which makes detailed measurements of the magnetoelectric properties difficult. We thus select Co$^{2+}$ ($S$=3/2) -based compound instead of Cu-based one in order to raise the magnetic transition temperature.
The crystal structures of Ca$_2$CoSi$_2$O$_7$ (Sr$_2$CoSi$_2$O$_7$) and Ba$_2$CoSi$_2$O$_7$ are shown in Figs.\@ 1(a) and (b), respectively \cite{CaCo, BaCo}.
As seen from Figs.\@ 1(a) and (b), both of them have a two-dimensional structure in which SiO$_4$ and CoO$_4$ tetrahedra are connected through the corners, but the arrangement of SiO$_4$ and CoO$_4$ tetrahedra is different from each other. The crystal structure consists of an alternate stack of the two-dimensional CoSi$_2$O$_7$ and Ca$^{2+}$ (Ba$^{2+}$) layers along the $c$ ($b$) axis. However, little is known about the physical properties of Ca$_2$CoSi$_2$O$_7$ and Ba$_2$CoSi$_2$O$_7$, and Sr$_2$CoSi$_2$O$_7$ has not been investigated so far. In this work, we have investigated the magnetic and dielectric properties of  $A_2$CoSi$_2$O$_7$.

\section{Experiment}
Single crystalline samples were grown by the floating zone method. Sample characterization was  performed by powder X-ray diffraction measurements at room temperature. We confirmed that the obtained crystals are of single phase and that Sr$_2$CoSi$_2$O$_7$ has the same crystal structure as Ca$_2$CoSi$_2$O$_7$. All the specimens used in this study were cut along the crystallographic principal axes into a rectangular shape by means of X-ray back-reflection Laue technique.  The magnetic properties were measured using a commercial apparatus (Quantum Design, Physical Property Measurement System (PPMS)). The dielectric constant was measured at 100 kHz using an {\it LCR} meter (Agilent, 4284A).

\section{Results and Discussion}

\begin{figure}[tb]
\begin{center}
\includegraphics[width=0.98 \textwidth,clip]{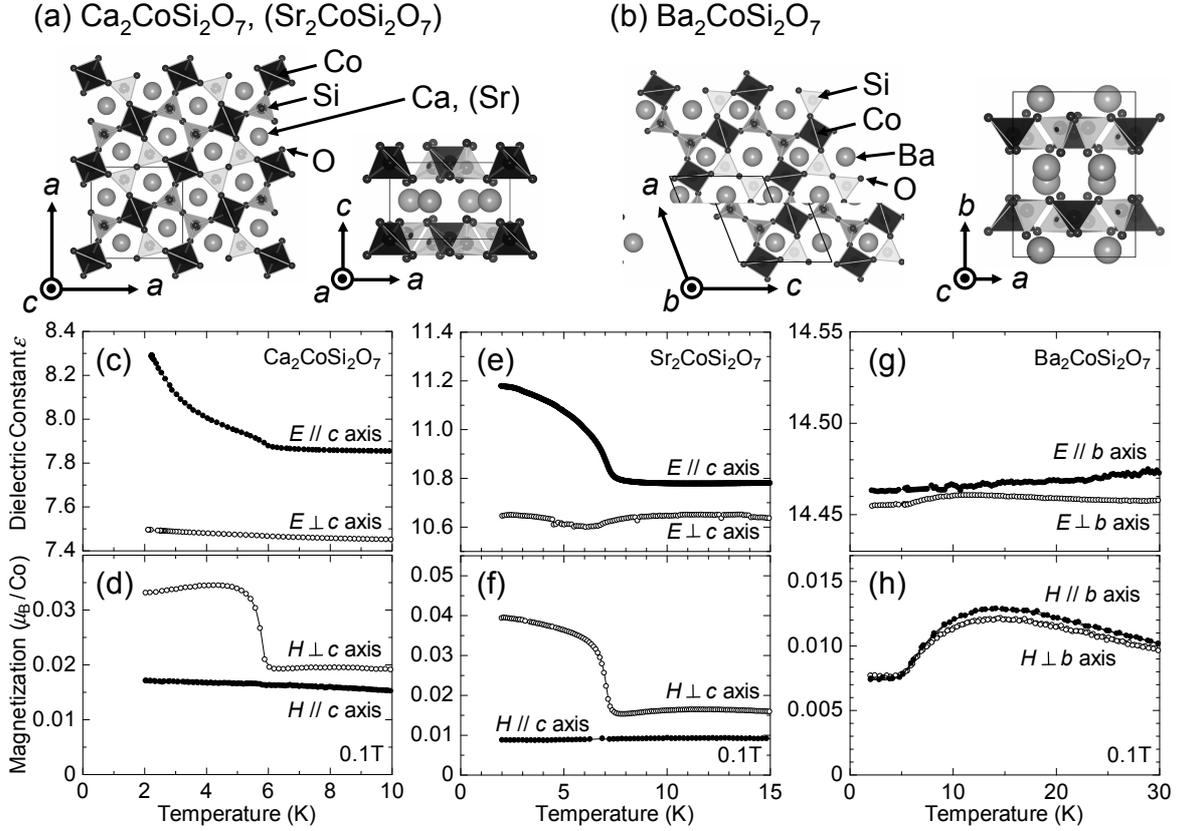}
\end{center}
\caption{The crystal structures of Ca$_2$CoSi$_2$O$_7$ (Sr$_2$CoSi$_2$O$_7$) (tetragonal, space-group $P\overline{4}2_1 m$) (a) and Ba$_2$CoSi$_2$O$_7$ (monoclinic, space-group $C2/c$) (b). Temperature dependence of dielectric constant (c), (e), (g) and magnetization (d), (f), (h) of $A_2$CoSi$_2$O$_7$ crystals. The magnetization after zero-field-cooling and dielectric constant were measured in warming scan.}
\label{f1}
\end{figure}

\begin{figure}[tb]
\begin{center}
\includegraphics[width=0.98 \textwidth,clip]{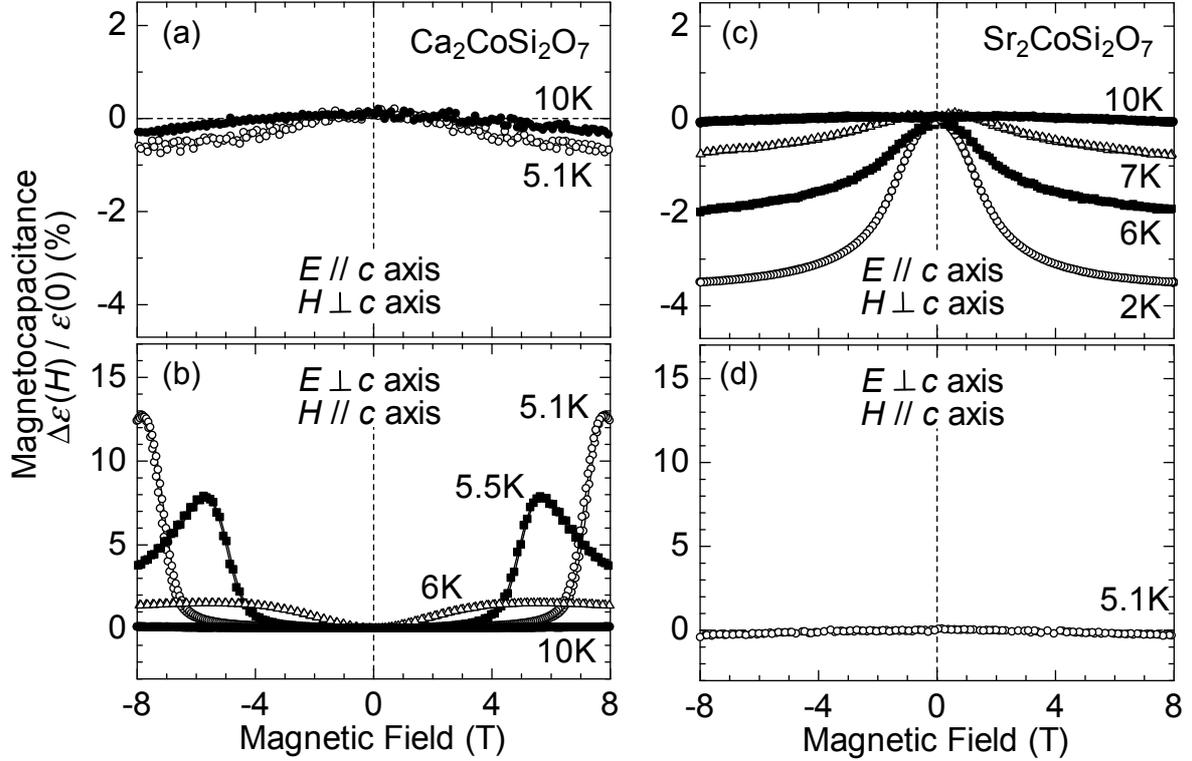}
\end{center}
\caption{Magnetic field dependence of magnetocapacitance of Ca$_2$CoSi$_2$O$_7$ (a), (b) and Sr$_2$CoSi$_2$O$_7$ (c), (d) with different measurement configurations at several fixed temperatures.}
\label{f2}
\end{figure}

Figure 1 shows the temperature dependence of the dielectric constant and magnetization of $A_2$CoSi$_2$O$_7$. The magnetization ($M_{\perp}$) of Ca$_2$CoSi$_2$O$_7$ shows a jump at 5.7 K when applying magnetic fields perpendicular to the $c$ axis (Fig.\@ 1(d)), indicating that a weak ferromagnetic (WF) transition occurs at the temperature. On the other hand, no anomaly is found in the magnetization ($M_{\parallel}$) measured with external magnetic fields parallel to the $c$ axis. The dielectric constant perpendicular to the $c$ axis ($\varepsilon _{\perp}$) does not show any anomaly, while the dielectric constant parallel to the $c$ axis ($\varepsilon _{\parallel}$) does a slight increase below the WF transition temperature ($T_{\rm WF}$) (Fig.\@ 1(c)). The simultaneous change of the $M_{\perp}$ and $\varepsilon _{\parallel}$ at 5.7 K implies that the magnetism is coupled with the dielectricity in Ca$_2$CoSi$_2$O$_7$. As seen from Figs.\@ 1 (e) and (f), the magnetic and dielectric properties of Sr$_2$CoSi$_2$O$_7$ are similar to those of Ca$_2$CoSi$_2$O$_7$. In Sr$_2$CoSi$_2$O$_7$, the $T_{\rm WF}$ shifts to slightly higher temperature of 7 K, below which the $\varepsilon _{\parallel}$ shows an abrupt increase compared with that of Ca$_2$CoSi$_2$O$_7$. On the other hand, the magnetic and dielectric properties of Ba$_2$CoSi$_2$O$_7$ are quite different from those of Ca$_2$CoSi$_2$O$_7$ and Sr$_2$CoSi$_2$O$_7$. 
The magnetization of Ba$_2$CoSi$_2$O$_7$ shows an anomaly at 5 K (Fig.\@ 1(h)), suggesting that some magnetic transition occurs, while the $\varepsilon _{\perp}$ shows little change at the temperature (Fig.\@ 1(g)). These results suggest that a coupling between the magnetization and dielectric constant is very weak in Ba$_2$CoSi$_2$O$_7$.

In Fig.\@ 2, we show the magnetic field dependence of the magnetocapacitance ($\Delta \varepsilon (H)/\varepsilon (0) \equiv [\varepsilon (H)-\varepsilon (0)]/\varepsilon (0)$) at several fixed temperatures. 
Below the $T_{\rm WF}$, the $\varepsilon _{\perp}$ of Ca$_2$CoSi$_2$O$_7$ strongly depends on magnetic fields parallel to the $c$ axis, and the large positive magnetocapacitance is observed (Fig.\@ 2(b)).\@ With decreasing temperature, the peak position of the magnetocapacitance curves shifts to higher magnetic fields of 5 T (5.5 K) and 8 T (5.1 K), and the magnetocapacitance effect is further enhanced; $\Delta \varepsilon (H)/\varepsilon (0)$ reaches 13 \% at 5.1 K\@. The magnetocapacitance of Ca$_2$CoSi$_2$O$_7$ is relatively large compared with those of other recently discovered multiferroic materials (TbMnO$_3$: 10 \% \cite{TbMn}, MnWO$_4$: 4 \% \cite{MnW}, LiCu$_2$O$_2$: 0.4 \% \cite{LiCu}). The observed large magnetocapacitance effect provides clear evidence for a strong coupling between the magnetism and dielectricity in Ca$_2$CoSi$_2$O$_7$. Although Ca$_2$CoSi$_2$O$_7$ does not show spontaneous ferroelectric polarization in the absence of magnetic fields (not shown), applying magnetic fields induces electric polarization below the $T_{\rm WF}$. 
This is so-called "magnetic-field-induced pyroelectricity" \cite{new}, which has not been reported so far in other multiferroic materials to our knowledge.
The $\varepsilon _{\parallel}$ of Ca$_2$CoSi$_2$O$_7$ is slightly suppressed by applying magnetic fields perpendicular to the $c$ axis (Fig.\@ 2(a)), i.e., the negative magnetocapacitance is found.

In Sr$_2$CoSi$_2$O$_7$, the $\varepsilon _{\parallel}$ depends on magnetic fields perpendicular to the $c$ axis.
The relatively large negative magnetocapacitance is observed below the $T_{\rm WF}$ (Fig.\@ 2(c)); $\mid\Delta \varepsilon (H)/\varepsilon (0)\mid$ reaches 3.5 \% at 2 K. Compared with the other multiferroic materials without magnetic-field-induced polarization, the magnetocapacitance of Sr$_2$CoSi$_2$O$_7$ is relatively large (BiMnO$_3$: 0.5 \% \cite{BiMn}, TeCuO$_3$: 1.0 \% \cite{TeCu}, BaCo$_2$Si$_2$O$_7$: 0.2 \% \cite{BaCo2}). The negative magnetocapacitance is ascribed to suppression of the $\varepsilon _{\parallel}$ by applying magnetic fields perpendicular to the $c$ axis below the $T_{\rm WF}$. In contrast to the case of Ca$_2$CoSi$_2$O$_7$,
the $\varepsilon _{\perp}$ of Sr$_2$CoSi$_2$O$_7$ is independent of magnetic fields parallel to the $c$ axis, and magnetic-field-induced pyroelectricity does not show up.

In Ba$_2$CoSi$_2$O$_7$, the dielectric constant is insensitive to magnetic fields (not shown), and electric polarization does not appear. This means that the correlation between the magnetism and dielectricity is almost negligible in Ba$_2$CoSi$_2$O$_7$. 
The difference among the magnetoelectric behaviors of Ca$_2$CoSi$_2$O$_7$, Sr$_2$CoSi$_2$O$_7$ and Ba$_2$CoSi$_2$O$_7$ is probably due to the difference in their two-dimensional networks of CoO$_4$ and SiO$_4$ tetrahedra. Therefore, further information on their crystal and magnetic structures are required for a full understanding of the mechanism of the large magnetocapacitance effect of $A_2$CoSi$_2$O$_7$. Synchrotron X-ray and neutron diffraction measurements are now in progress.

\section{Conclusion}
In summary, we have investigated the magnetic and dielectric properties of $A_2$CoSi$_2$O$_7$ ($A$=Ca, Sr, and Ba) and have observed the large magnetocapacitance effect in Ca$_2$CoSi$_2$O$_7$ and Sr$_2$CoSi$_2$O$_7$ crystals. The large magnetocapacitance effect indicates a strong coupling between the magnetism and dielectricity in Ca$_2$CoSi$_2$O$_7$ and Sr$_2$CoSi$_2$O$_7$. In contrast, Ba$_2$CoSi$_2$O$_7$ hardly shows the magnetocapacitance, indicating that a coupling between the magnetism and dielectricity is almost negligible.
The arrangement of CoO$_4$ and SiO$_4$ tetrahedra is significant for the large magnetocapacitance of $A_2$CoSi$_2$O$_7$.

\section*{Acknowledgment}
This work was supported by Grant-in-Aid for scientific research (C) from the Japan Society for Promotion of Science.
\\


\begin{thebibliography}{9}
\bibitem{new} The detailed results will be reported elsewhere.
\end{thebibliography}
\end{document}